\documentclass[12pt]{article}
\usepackage{amsmath,amssymb,amsfonts,latexsym,epsfig,times,color}

\begin{document}

\title{A Scaling Law for Quark Masses}
\author{Harald Fritzsch {\small and } Alp Deniz \"Ozer   \\
{\small Ludwig Maximillians University,   }  \\
{\small Sektion Physik, Theresienstr. 37, 80333  M\"unchen
Germany} }
\date{}

\maketitle

\abstract{We show that the observed quark masses seem to be
consistent with a simple scaling law. Due to the precise values
of the heavy quarks we are able to calculate the quark masses in
the light quark sector.  We discuss a possible value for the
strange quark mass. We show that the u-type quark masses obey the
scaling law  very well.}\\

%\section*{Introduction}

The masses of the quarks are important parameters of the
Standard Model but thus far, have remained unexplained. In the
Standard Model they are generated by the coupling of the quark
fields to the hypothetical scalar boson, which breaks the $SU (2)
\times U(1)$ symmetry.

The observed quark masses show a remarkable pattern. The $u$ and
 $d$ masses are relatively light, a few MeV in mass, the $s$
and $c$ masses are in the region of $100 \dots 1200$ MeV, while
the $b$ and $t$ masses are heavy. The t-mass of about 170 GeV is
the only quark mass, which is of the same order as the energy
scale describing the violation of the $SU(2)$ gauge symmetry.

In this paper we would like to argue that the quark masses might
follow a simple scaling law:
\begin{equation}\label{sl}
\begin{split}
 &\frac{m_t}{m_c}=\frac{m_c}{m_u}    \\
 & \\
 &\frac{m_b}{m_s}=\frac{m_s}{m_d}    \\
\end{split}
\end{equation}
Let us first demonstrate that the observed masses of the quarks
might actually be  consistent with the simple scaling laws. Of
course, such laws make sense only, if the quark masses are all
renormalized at the same energy scale. For the u-type-quarks we
choose the central value of $m_t=174.3 \pm 5.1 $ GeV as a useful
scale. We choose for the c-quark mass $m_{c}(m_c)=1.27 \pm 0.05 $
GeV given in~\cite{Gasser:1982ap}, which rescales to $m_c{(m_t)}=0.62 \pm 0.03
$ GeV, using the QCD renormalization
group~\cite{Muta} with $\Lambda=211^{+34}_{-30}$ MeV for five
flavors ~\cite{Bethke:2002rv}. Then one finds
\begin{equation}
\begin{split}
 &\frac{m_t}{m_c} =  260 \dots 304 \\
\end{split}
\end{equation}
The u-mass $m_{u}$ is given as $m_{u}(1\, \text{GeV})=5.1 \pm 0.9
$ MeV~\cite{Leutwyler:1996eq}. Using the QCD renormalization group with
$\Lambda=211^{+34}_{-30}$ MeV for five flavors,  one has
$m_{u}(m_{t})=2.28 \pm 0.41 $ MeV. Then we obtain:
\begin{equation}
\begin{split}
 &\frac{m_c}{m_u} =220\dots348 \\
\end{split}
\end{equation}
Both ratios are of the same order of magnitude. If they are set to
be equal, we find for the central value of the mass ratios:
\begin{equation}
\begin{split}
 &\frac{m_t}{m_c}=\frac{m_c}{m_u} = 281  \\
\end{split}
\end{equation}
Thus we obtain
\begin{equation}\label{uctmasses}
\begin{split}
 m_{c}(m_c)& = 1.27 \, \, \text{GeV} \\
 m_{u}(2\,\text{GeV})& = 3.94 \, \, \text{MeV} \\
 \end{split}\ \ \ \ \
 \begin{split}
 m_{c}(m_{t}) & = 0.62 \, \, \text{GeV} \\
 m_{u}(m_{t}) & = 2.21 \, \, \text{MeV} \\
 \end{split}
\end{equation}
Here the charm quark mass $ m_{c}(m_c)=1.27$ GeV  is
consistent with $m_c(m_c)$ $=1.23 \pm 0.09$ GeV calculated  from
QCD sum rules in the charmonium system given in ~\cite{Eidemuller:2000rc}.
Indeed  the top and charm quark are among the "heavy quarks", and
their masses are known within small error bars~\cite{Gasser:1982ap}.
Therefore the scaling law prediction for the up-quark mass is
quite definitive. The error in the mass of u-quark stems from the
error in the charm and top quark masses and depends also on the
error in $\Lambda_{QCD}$~\cite{Bethke:2002rv}.

 %%%%%%%%%%%%%%%%%%%%%%%%%%%%%%%%%%%%%
 %%%%%%%%%%%%%%%%%%%%%%%%%%%%%%%%%%%%%
 %%%%%%%%%%%%%%%%%%%%%%%%%%%%%%%%%%%%%
 %%%%%%%%%%%%%%%%%%%%%%%%%%%%%%%%%%%%%
 %%%%%%%%%%%DDDDDDDDDDDDDDDD%%%%%%%%%%
 %%%%%%%%%%%%%%%%%%%%%%%%%%%%%%%%%%%%%
 %%%%%%%%%%%%%%%%%%%%%%%%%%%%%%%%%%%%%
 %%%%%%%%%%%%%%%%%%%%%%%%%%%%%%%%%%%%%
The same can be done for the d-type quarks. But among the d-type
quarks $only$ the bottom quark is a "heavy quark" and has a
relatively well known mass. On the contrary the error in the
strange quark mass is rather high. Consequently the ratio
$m_b/m_s$ will contain large uncertainties, and the scaling law
prediction for the down quark mass will not be definitive. However
using  big error bars for the strange quark mass, it is still
possible to have a consistent scaling law.

The scaling of the d-type quarks will be done at $2$ GeV. For the
bottom quark we choose $m_b(m_b)=4.25 \pm 0.10$ \cite{Gasser:1982ap}
which rescales to $m_b(\text{2 GeV})=5.02 \pm 0.14 $ GeV by
using the QCD renormalization group with the current value
$\Lambda=294^{+42}_{-38}$ MeV, for 4 flavors given in "$\alpha_s$
2002"~\cite{Bethke:2002rv}.  We chose for the strange quark mass $m_s(1
\,\text{GeV}) = 175 \pm 55 $ MeV \cite{Gasser:1982ap}, which rescales to
$m_s=134 \pm 42 $ MeV  at $2$ GeV by using the QCD
renormalization group with the current value
$\Lambda=336^{+42}_{-38}$ MeV, for 3 flavors given in
~\cite{Bethke:2002rv}. The down quark mass is chosen as $m_d(  \text{1
GeV})=9.3 \pm 1.4$ MeV \cite{Leutwyler:1996eq} which rescales to $m_d(
\text{2 GeV}) = 7.1 \pm 1.1 $ MeV. Then we obtain :
\begin{equation}
\begin{split}
  \frac{m_b}{m_s}& = 28 \dots 56 \\
\end{split}\ \ \ \ \ \ \ \ \
\begin{split}
  \frac{m_s}{m_d}& = 11 \dots 29  \\
\end{split}
\end{equation}
which are again of the same order of magnitude. We can set them equal
and find for the central values of the mass ratios:
\begin{equation}
\begin{split}
 &\frac{m_b}{m_s}=\frac{m_s}{m_d}=  28  \\
\end{split}
\end{equation}
Requiring the mass ratio to be the same, we have :
\begin{equation}\label{d-type}
\begin{split}
 m_d(\text{2 Gev})& =  6.51 \, \, \text{MeV} \\
 m_s(\text{2 Gev})& =  180    \, \, \text{MeV} \\
\end{split}\ \ \ \ \ \ \ \
\begin{split}
 m_d(m_b)& =   5.60  \, \, \text{MeV} \\
 m_s(m_b)& =   155 \, \, \text{MeV} \\
\end{split}
\end{equation}
Here  $ m_{b}(m_b)=4.25$ GeV is consistent with an independent
analysis giving  $m_b(m_b)$ $=4.20 \pm 0.09$ GeV, derived from
low-n sum rules  in ~\cite{Corcella:2003xx}.

%\section*{Conclusion}

Thus we conclude that the observed quark masses seem to be consistent with the simple
scaling laws.  The best values for the ratio of the u-type quark
masses is $281$ , for the d-type masses it is $28$. The quark
masses that we found through the scaling law are consistent with
current values of quark masses.

The strange quark mass from an analysis from the observed spectrum
of $\tau$ decay ~\cite{Barate:1999hj} predicts $m_{s}=170^{+44}_{-55}$ at
2 GeV and is consistent with the scaling law.

Of course in the Standard Model there is no reason for any scaling
law. Such a reason can only be given in specific theories beyond
the Standard Model. In this paper we do not wish to speculate
about such reasons. However the fact that the scaling law
discussed by us could be either approximately or perhaps even
exactly true seems interesting and should be investigated further.

This Paper is dedicated to  the memory of Prof. D. Tadic,
who has contributed much to the investigation of the quark mass problem.

\end{document}